# Anharmonic Vibrational Spectrum and Experimental Matrix Isolation Study of Thioformic Acid Conformers – Potential Candidates for Molecular Cloud and Solar System Observations?

Antti Lignell[1,3,†,*], Irina Osadchuk[1,2,†], Markku Räsänen[3], and Jan Lundell[1,*]

Abstract

Thioformic acid (TFA) is the sulfur analog of formic acid, the simplest organic acid. It has three analogues HCOSH, HCSOH, and HCSSH, each of them having two rotational isomeric (rotameric) forms: trans and cis where the trans form is energetically more stable. In this article, we study computational energetics and anharmonic vibrational spectrum of TFA including overtone and combination vibrations. We also studied experimental photoisomerization and photodecomposition channels of HCOSH molecules with different wavelengths. We suggest that TFA is a potential sulfur containing candidate molecule for interstellar and planetary observations and discuss these in a light of different radiation environments in space. More generally, we discuss that infrared radiation driven photo-isomerization reactions may be a common phenomenon in such environments and can affect the chemical reaction pathways of organic and other interstellar molecules.

[1] Department of Chemistry, University of Jyväskylä, P.O. Box 35, FI-40014 University of Jyväskylä, Finland
[2] Department of Chemistry, Tallinn University of Technology, Akadeemia tee 15, 12618 Tallinn, Estonia
[3] Department of Chemistry, University of Helsinki, P.O. Box 64, FI-00014 University of Helsinki, Finland
† Equal contribution

* Corresponding authors: AL Tel: +358505911666; lignell@gmail.com, JL Tel: +358-407445270; E-mail: jan.c.lundell@jyu.fi

# 1. Introduction

Sulfur is the second lightest element in the group 16 (oxygen family) and it is ubiquitously found in Earth (5$^{th}$ most common element by mass), elsewhere in the Solar System, and the Universe (Caffau et al. 2005; Morgan & Anders 1980; Simionescu et al. 2015; Vidal et al. 2017). It can be found for example in elemental form ($S_8$), as sulfate minerals in planetary bodies, gases (e.g. $H_2S$ and $SO_2$), and in complex organic molecules. It frequently forms chemical compounds similar to their more well-known oxygen analogs, such as $H_2S$ (water), $CS_2$ (carbon dioxide), thiols (alcohols), thioketones, thioaldehydes, and sulfides (thioethers). Sulfur is an essential element in all living cells as it can be found in two of the naturally occurring amino acids (methionine and cysteine), proteins and enzymes, vitamins (biothine and thiamine), co-factors (glutathione), and hormones such as insulin and oxytocin (Francioso et al. 2020). Sulfur chemistry is an energy source for many life forms that live in complete darkness, e.g. around the hydrothermal vents on the seafloor (Baross & Hoffman 1985; Martin et al. 2008; Sievert et al. 2008). This type of energy source has been suggested to exist elsewhere in the solar system icy bodies, such as in the subsurface ocean of Europa that may harbor conditions suitable for life (Lowell & DuBose 2005; Pappalardo et al. 1999). Sulfur also has an important role in atmospheric chemistry and climate change since sulfur-containing aerosols are common cloud condensation nuclei and many atmospherically important reactions such as ozone depletion can take place on their surface (Portmann et al. 1996; Sipila et al. 2010; Solomon et al. 1996; Tie & Brasseur 1995).

Sulfur compounds are common in planetary bodies and interstellar space. Most notably in Io, the innermost Galilean moon of Jupiter, where the majority of the surface is covered with sulfur compounds that are partially responsible for the deep yellow, orange, red, brown, and green coloration of the moon (Carlson et al. 2007; Geissler et al. 1999b). The huge tidal forces due to the Io's close proximity to Jupiter leads to tidal heating and subsequent vast volcanic activity on its surface, Sulfur-rich eruptions and "blooms" have been observed on Io. The most of that material is deposited onto the surface of the moon but also Jupiter's magnetosphere sweeps a significant amount (~1 ton per second) of this material into the Jovian system and beyond (Geissler et al. 1999a; Schneider & Bagenal 2007).

The surface of Io has been studied with both Earth-based and remote sensing solar reflection spectroscopy, $SO_2$ being the dominant species but other sulfur molecules such as $H_2S$ and $Cl_2SO_2$ have also been characterized (Carlson et al. 1992; Lopes-Gautier et al. 2000; Tosi et al. 2020). Interestingly, sulfur compounds have been found from the atmosphere of exoplanet 51 Eri b suggesting that these compounds may also play a critical role in chemistry and haze formation in exoplanets throughout the Universe (Zahnle et al. 2016).

Interstellar space and circumstellar discs provide conditions for a surprisingly complex chemistry and many species from simple radicals to more complex organic molecules have been found (Oberg 2016). Various spectroscopy techniques, both absorption and emission, have been used to identify these species. Majority of the species are observed from their rotational microwave absorption/emission, but also vibrational spectroscopy at infrared region has been used for their characterization. The cold vacuum-like interstellar conditions stabilize many radicals and other short-lived systems increasing the number of more "exotic" chemical species in the list. Many sulfur compounds have been found from simple inorganic compounds (e.g. $SO_2$, CS, OCS) to complex organics such as thioformaldehyde and methyl mercaptan (Jefferts et al. 1971; Lamberts 2018; Minh, Irvine, & Brewer 1991; Penzias et al. 1971; Snyder et al. 1975).

The simplest carboxylic acid, formic acid (FA), has been observed in interstellar medium already since 1971 from Sagittarius B2 molecular cloud (Zuckerman, Ball, & Gottlieb 1971). It has two rotational isomers: trans-HCOOH and cis-HCOOH where the trans-conformer is energetically more stable (1365 $cm^{-1}$ energy difference)(Hocking 1976) and it has a rotational barrier ranging ~4000-4500$cm^{-1}$.(Hirao 2008; Marushkevich et al. 2010; Pereira et al. 2014) Energetically less favorable cis-conformer could be enriched by exciting the molecule over its rotational barrier e.g. with infrared radiation and stabilized in a low temperature matrix (Macoas et al. 2003; Marushkevich et al. 2006; Marushkevich, Khriachtchev, & Rasanen 2007; Pettersson et al. 1997). Interestingly, Cuadrado et al. recently observed cis-FA in Orion Bar photodissociation region of the Orion Nebula and they suggested that it is formed via UV-mediated conformational change (photoswitching) in the molecular cloud (Cuadrado et al. 2016). According to our best knowledge, this was the first observation of conformational change in interstellar space and is an important opening of conformational chemistry outside Earth. Potentially many other organic molecules could exhibit these conformational changes and enrichment of higher energy rotamers under similar conditions.

In this article we present the computational anharmonic spectra of both cis and trans conformers of mono- and di-thioformic acids (TFA) together with their energetics. The fundamental, overtone and combination vibrations are obtained at DFT hybrid functional B3LYP. Thioformic acid is the smallest homologue of the thiocarboxylic acids (Figure 1) and it is the less stable sulfur analog to formic acid. It has been studied computationally and theoretically for hydrogen bonding, proton exchange, and conformational changes (Delaere, Raspoet, & Nguyen 1999; Huang et al. 1998; Jemmis, Giju, & Leszczynski 1997; Kaur, Sharma, & Aulakh 2011; Kaur & Vikas 2014a, 2014b, 2015). We combined the relevant experimental gas-phase infrared and matrix isolation data for mono- and dithioformic acids (Dellavedova 1991; Winnewisser & Hocking 1980). Also, we present experimental photoswitching of TFA (HCOSH) in a low temperature matrix and its photodecomposition pathways under different photon environments that are relevant to astrophysical conditions.

## 2. Methods

All computations were performed with Gaussian16 program package. We used B3LYP method with aug-cc-pVTZ basis sets. (Becke 1988; Kendall, Dunning, & Harrison 1992; Lee, Yang, & Parr 1988). Anharmonic vibrational frequencies and intensities were simulated using Barone's method implemented in Gaussian16 (Bloino & Barone 2012). Zero-point energy correction was implemented with harmonic vibrational energy correction.

TFA was prepared according to the method by Engler and Gattow and modified by Hocking and Winnewisser (Engler & Gattow 1972; Hocking & Winnewisser 1976c). The raw product was purified by trap-to-trap distillations. The typical matrix-to-sample ratio was 500-1000 and the matrix gases were obtained from AGA (Ar, 99.998%) and Matheson (Xe, 99.997%), and the matrix data has been obtained at 15K temperature. The distinctive matrix deposition temperatures were 20K for argon and 28K for xenon. The infrared spectra were recorded with a Nicolet 60SX FTIR spectrometer using a Ge/KBr beam splitter and MCT detector for mid-IR and a 6μm Mylar beam splitter and DTGS/PE detector for far-IR. The spectral resolution was 0.25cm$^{-1}$ in mid-IR and 1cm$^{-1}$ in far-IR. Infrared excitations were performed by using a Globar radiation from the spectrometer with various interference filters and for UV-irradiations we used a 150W high-pressure mercury arc lamp with water filter and different cut-off filters. The idea of using noble-gases (argon and xenon) as a matrix material is to isolate individual molecules into inert media and thus simulate interstellar condition where molecules are well separated from each other. This is a powerful way to study reactive and otherwise unstable species as a trapped form in an inactive medium.(Khriachtchev 2011)

## 3. Results and discussion

We have collected the experimental and computational vibrational frequency data of TFA molecules in Table 1. In addition to fundamental vibrations, we present overtone and combination vibration bands for the most prominent absorptions. Our computational approach gives quite reliable estimates where the average error compared to known gas phase values in fundamental vibrations is ~1.5%. This is in line with the high-level vibrational energy calculations for formic acid (Tew & Mizukami 2016). We also calculated rotational constants of these molecules and they are presented in Table 2 together with the existing experimental data. The rotational constants require highly accurate molecular structures and our computational method was optimized for anharmonic vibrational energy calculations. These values are not accurate enough for direct astronomical observations. We present them to shed some light on the accuracy of our computational method to predict these values.

The structures of three different TFAs and their conformers are presented in Figure 1 and the coordinates in the Appendix. The coordinate file is available for download in text format (.xyz) as a supplementary material and it represents the positions of atoms in individual TFA molecules in Cartesian coordinates. The most notable difference in OH/SH and CS/CO bonds is the longer bond length for sulfur analogues compared to oxygen containing species. The energetics between cis/trans isomers and the transition states are presented in Figure 2. HCOSH form of TFA has the lowest transition state barrier for the rotation of alcohol/thiol group (~3250cm$^{-1}$) whereas the HCSOH has ~1200cm$^{-1}$ higher barrier. It is notable that the energy of cis-conformer of HCOSH is only 213cm$^{-1}$ higher than trans-TFA giving the cis/trans ratio of .351 in NTP conditions according to Bolzmann distribution. This ratio is significantly changed in lower temperatures and is reduced to .062 at the mean surface temperature (110K) of Io and to almost zero (~$10^{-13}$) at the typical temperature (10K) of a molecular cloud. If the low temperatures in molecular clouds make the thermal interconversion to cis-TFA highly unlikely, the photon environment could drastically change the actual distribution between the cis/trans ratio. Interestingly, cis-form of formic acid (HCOOH) was recently observed in the molecular cloud where the cis/trans ratio of ~0.357 was observed (~$10^{-13}$ at the corresponding thermal equilibrium). This was attributed to the photoswitching between the conformers due to the local stellar UV-radiation (Cuadrado, et al. 2016). Recent finding of c-HCOOH/t-HCOOH ratio of 6% has been detected in star-forming regions Barnard 5 and L483 that is showing ~800 times higher amount of cis-HCOOH that would be expected.(Agundez et al. 2019; Taquet et al. 2017) Due to the lack of strong UV field in these systems, the authors suggest the chemical formation mechanism for the excess cis-HCOOH. In our opinion, IR-mediated transformation to cis-HCOOH would be possible under such conditions as IR radiation from a nearby stars/protostars have much deeper penetration into a circumstellar disc or molecular cloud, and it does not have sufficient energy to photochemically decompose these molecules into smaller fragments. Rayleigh scattering ($I_R$) is inversely dependent on the fourth power of a wavelength ($\lambda$) of electromagnetic radiation ($I_R \propto \frac{1}{\lambda^4}$) and this ensures much deeper penetration of longer wavelength photons such as IR. It is also notable that orange and red dwarf stars, that are together by far the most common star types in Milky Way, do emit a significant proportion of their photons in IR wavelengths and that could induce these IR mediated conformational changes in their circumstellar discs and star systems.

Formic acid has a rotational barrier height of ~4000-4500cm$^{-1}$ and the trans-form is 1365cm$^{-1}$ lower in energy compared to cis conformer (Hirao 2008; Hocking & Winnewisser 1976a; Marushkevich, et al. 2010; Pereira, et al. 2014). Because of the lower barrier height and much closer energies between TFA conformers, the possible interstellar TFA molecules are expected to have even larger distribution in the cis-form. This effect should be more prominent deeper in the molecular clouds where the lower wavelength light penetration is compromised. Due to the observation of formic acid rotamers in space, it would be likely that TFA could exist under similar conditions given the related characteristics of the molecules and the high abundance of sulfur in space. Interestingly, molecular conformers may have different chemical reaction pathways as shown e.g. in photochemical dissociation products of cis and trans formic acid (Khriachtchev et al. 2002; Martinez-Nunez et al. 2005). This potentially has a noticeable effect for the reaction pathways and production of organic and other species in molecular clouds. Strikingly, during the editorial process of this article Rodriguez-Almeida et al. reported the first observation of a thioformic acid in the interstellar medium (Rodriguez-Almeida et al. 2021). The location is Galactic center quiescent cloud G+0.693-0.027 where the kinetic temperature is between 70-150K(Zeng et al. 2018). According to Boltzmann distribution and our calculated energy difference ($213 cm^{-1}$) the cis-HCOSH should be present between ~1-12% of the total amount of HCOSH in G+0.693-0.027 towards observer. Rodriguez-Almeida *et al.* estimated that the upper limit for the column density of cis-HCOSH is $\leq 3 \times 10^{12}$ $cm^{-2}$ whereas the column density of trans-HCOSH is $1.6 \times 10^{13}$ $cm^{-2}$. That gives the upper limit for the cis-trans ratio of <19% in G+0.693-0.027 and this is in line with our theoretical estimate (1-12%). However, we encourage Rodriguez-Almeida *et al.* and others to have a closer look at the possibility of finding cis-HCOSH isomer from that source.

Our experiments in matrixes show that there are three major energy distribution channels for interstellar TFA molecule HCOSH and we present them in Figure 3 and a sample of the experimental spectra in Figure 4. It is notable that cis and trans isomers of HCOSH are well distinguishable via their $\nu5$ and $\nu7$ vibration modes. These observations are in agreement with the related study of $H_2S$-CO complex in an argon matrix (Lundell, Nordquist, & Rasanen 1997). IR irradiation below $3200 cm^{-1}$ did not lead to any significant changes. Broadband IR irradiation over $3200 cm^{-1}$ resulted to efficient trans→cis rotamerization and this threshold energy agrees well with our barrier height calculations ($3253 cm^{-1}$). The UV irradiation wavelength higher than 350nm mainly opened the trans→cis rotamerization channel. UV irradiation wavelength below 350nm photo-decomposes TFA into smaller fragments. At the 350-340nm range, the main product was OCS while the shorter wavelengths decomposed TFA to $H_2S$ and CO. These were monitored from their IR vibrational frequencies and they are collected into Table 4. Interestingly, OCS molecule itself has been linked as a mediator of polymer formation from simple amino acids suggesting its important role in prebiotic chemistry and the origin of life (Leman, Orgel, & Ghadiri 2004). The synthesis of TFA under astrophysical condition may involve hydrogenation of OCS species on dust grain surfaces and it can participate to interstellar reaction pathways of OCS and its potential role to astrobiology.(Palumbo, Geballe, & Tielens 1997) Another likely formation mechanism for interstellar TFA include HCO + SH radical-radical route.(Rodriguez-Almeida, et al. 2021)

None of the fundamental vibrations has sufficient energy to overcome the barrier without tunneling reaction and this is effectively demonstrated in the experiments that show the lack of trans→cis conversion with Globar radiation energy under the barrier height ($3200 cm^{-1}$). In the case of HCOSH molecule, the barriers height of $3250 cm^{-1}$ allows conversion to take place via the first $\nu_3$ overtone as well ($3420 cm^{-1}$ in the gas phase), that may be a dominant route due to the relatively strong absorption intensity. Also the first overtone of SH-stretching $\nu_2$ at ~$5000 cm^{-1}$ offers an effective channel to overcome the barrier. Other potential channels include $\nu_1$ overtone and $\nu_3+\nu_1$ combination vibrations. In principal, any overtone or combination vibration that has enough energy to overcome the barrier could be responsible for the conversion due to the effective internal vibrational relaxation and energy transport.(Tesar et al. 2013)

## 4. Conclusions

As conclusion, thioformic acid and its conformers are potential molecules for interstellar and solar system observations. We presented here experimental photoswitching between trans-cis HCOSH molecule and its photodecomposition routes with different wavelength radiation. We give computational energetics and anharmonic vibration at energy levels for fundamental, overtone, and combination vibrations of different TFAs and speculate the possible channels for the cis-trans isomerization of TFA in space. We also suggested that more generally, the rotational isomers of different molecules could be a very intriguing subject for observation of chemistry in space and the laboratory simulations of such systems could be useful for modeling these conditions.

## 5. Acknowledgments

We thank Eija Nordqvist and Janne Nieminen for their help with the matrix isolation experiments of this paper.

Figures

Figure 1. The computational minimum energy structures of three thioformic acids and their conformers. The colors are referring to the following elements: carbon (graphite), hydrogen (silver), oxygen (red), and sulfur (yellow). The more detailed structures are presented in the Appendix.

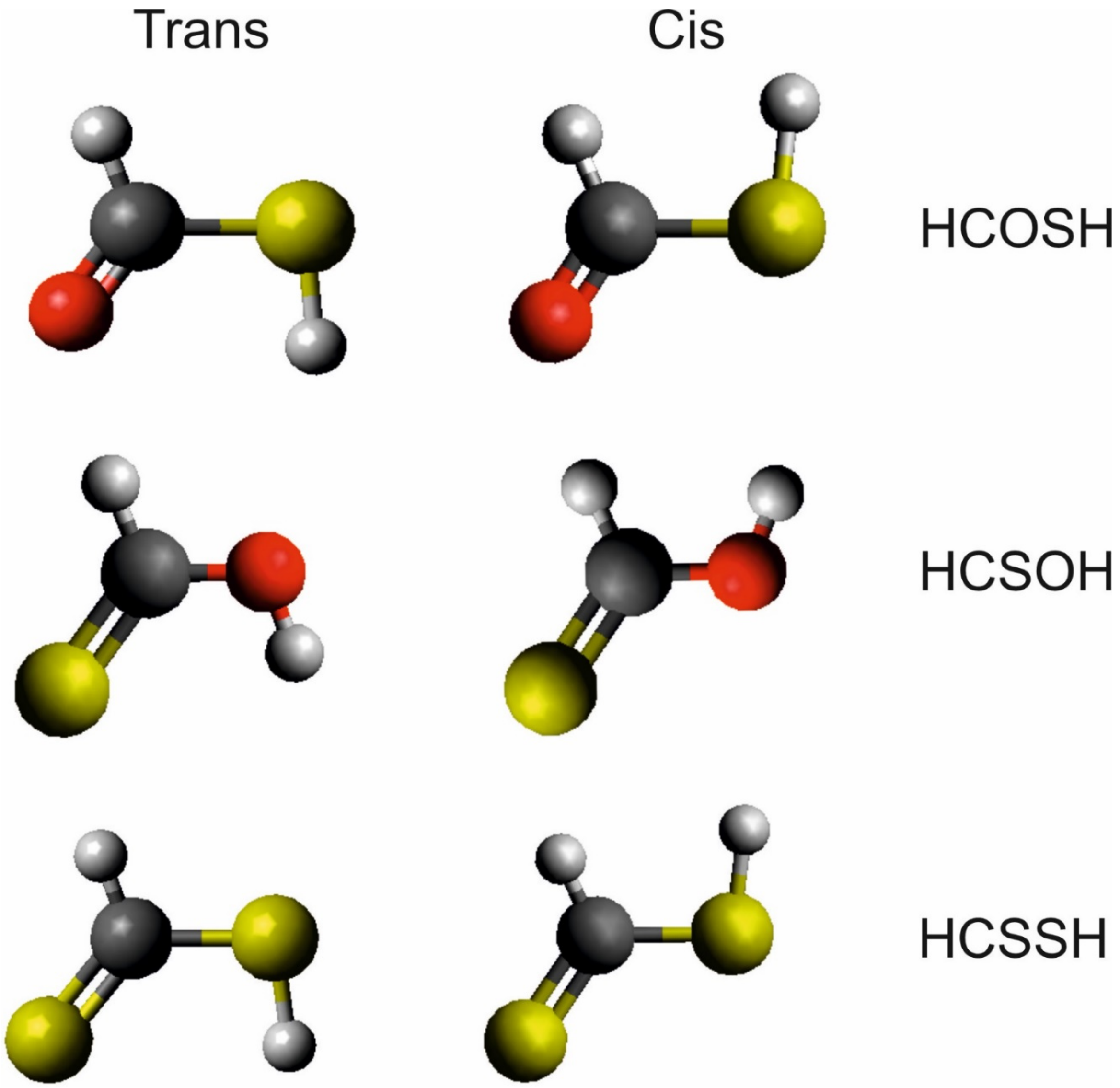

Figure 2. Computational vibrational zero-point energy corrected energies of thioformic acid conformers and their transition states with B3LYP/aug-cc-pVTZ level of theory. It is notable that HCSOH molecule has the highest transition and cis-conformer energies. The energies of trans-TFAs have been set to zero. The figure presents only the HCOSH molecule structure.

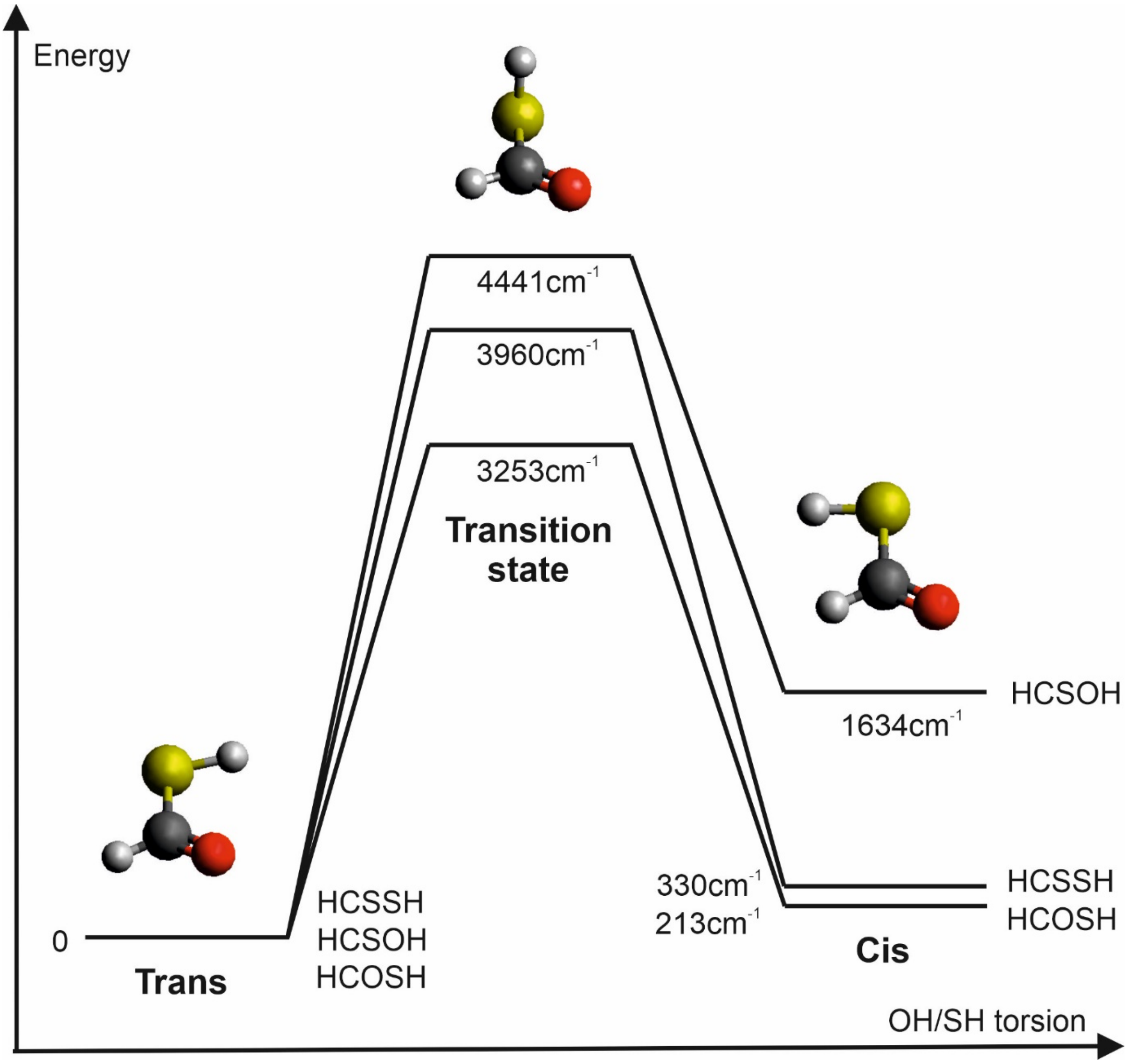

Figure 3. IR and UV induced changes in thioformic acid (HCOSH). IR irradiation with energy exceeding 3200 $cm^{-1}$ and the UV photolysis with >350nm leads primary to trans → cis conformational change in matrixes. UV irradiation with <350nm results into photodecomposition of TFA via two channels. The OCS channel is favored at the UV energy between 340-350nm whereas the $H_2S$ + CO channel is dominating with higher energies. The threshold for IR induced conformational change is in line with our transition state energy calculations, presented in Fig 2. The energy scheme presented in the lower panel show the potential astrophysical conditions (photon environment) where TFA conformers and their photoproducts could be stable.

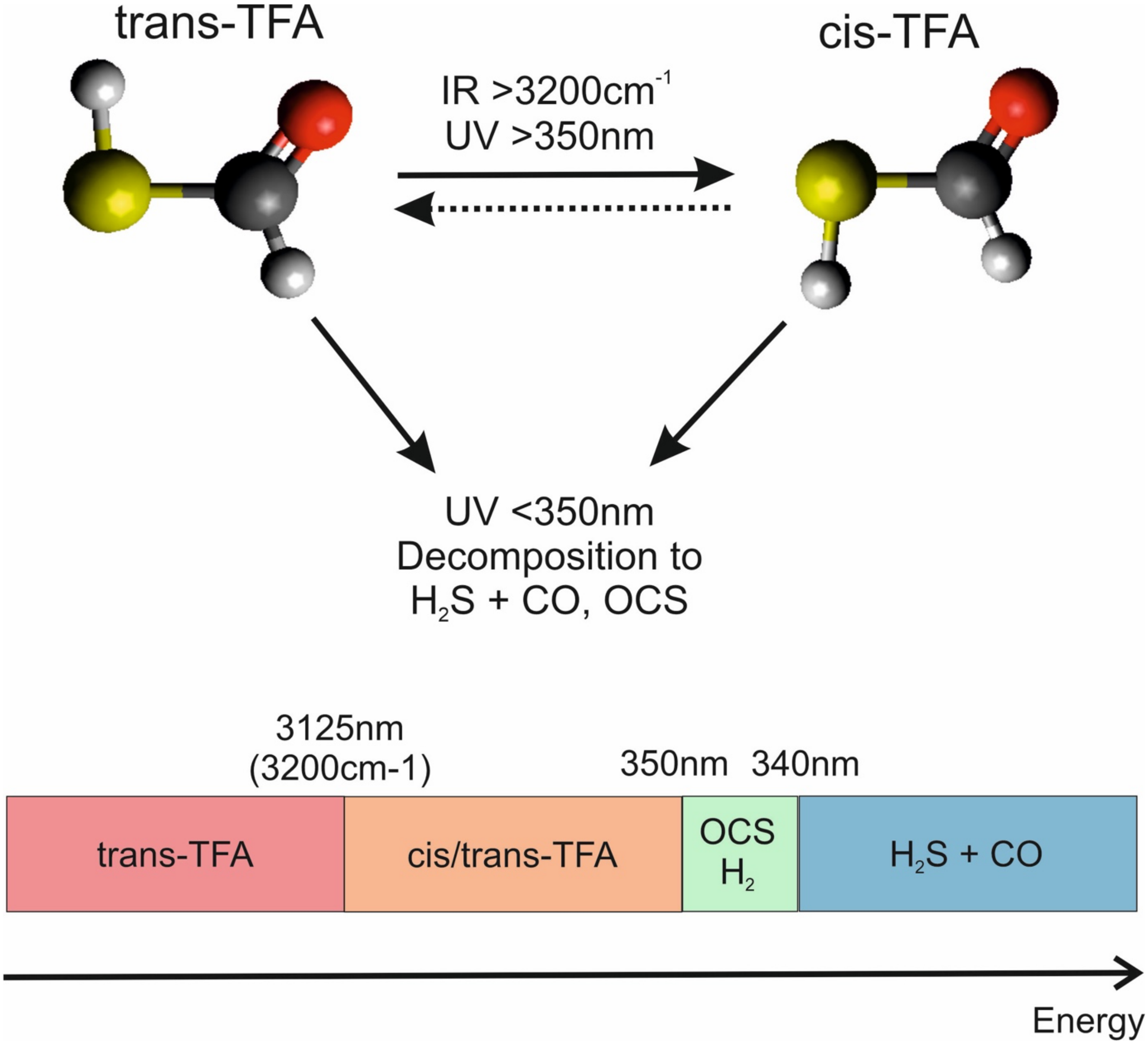

Figure 4. Infrared spectra of thioformic acid (HCOSH) in argon matrixes. Upper traces present the initial situation after deposition. Left panels present the ν5 CSH bending and ν6 CH wagging absorptions regions and the right panels ν7 CS stretching absorptions. The middle and lower panels show pure conformer absorptions obtained after IR-induced changes.

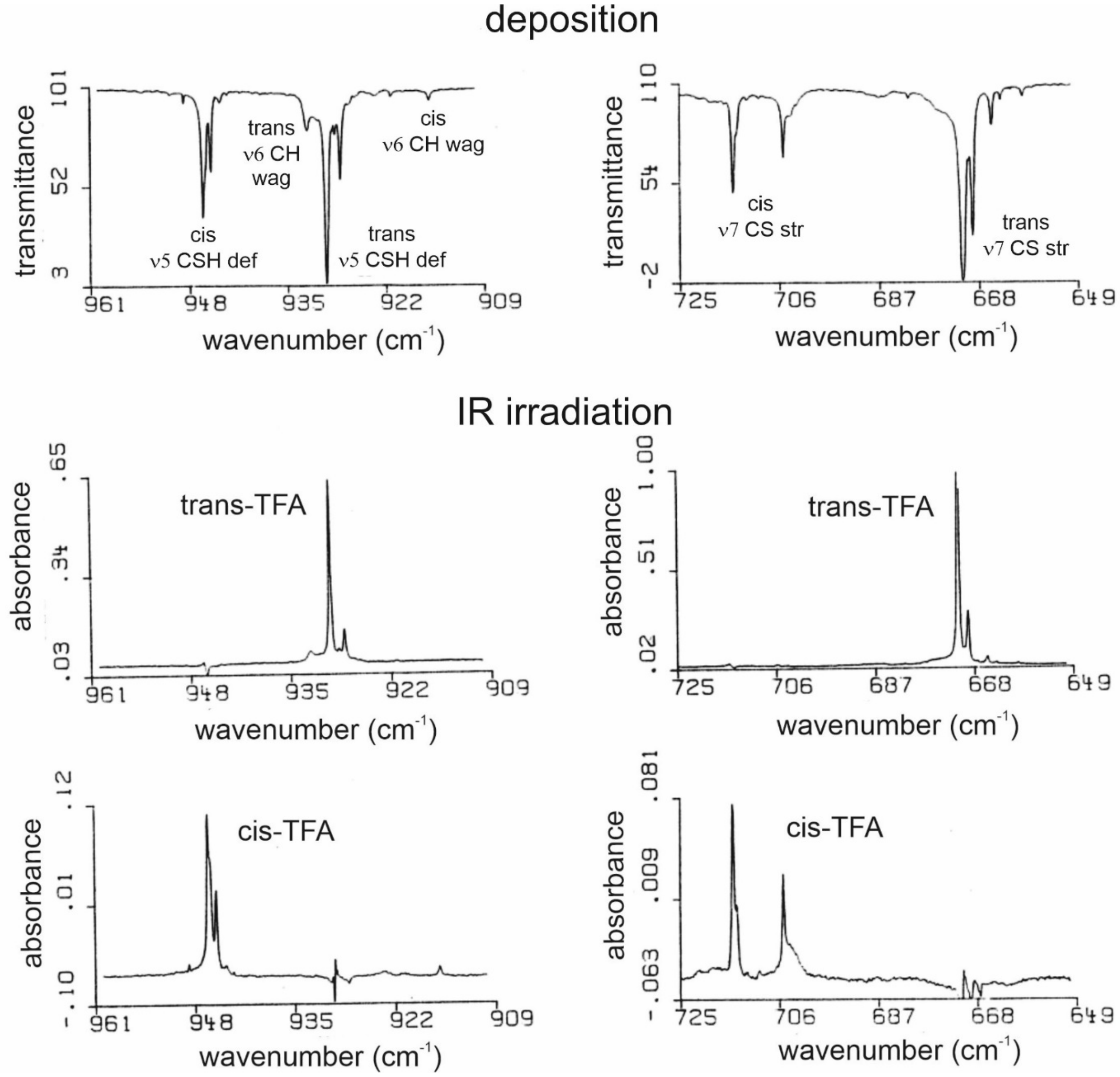

Table 1. Computational and experimental vibrational frequencies of three thioformic acid isomers. The computational level is B3LYP/aug-cc-pVTZ by using anharmonic potentials (see Methods) and we have included overtone and combination vibrations. The computational absorption intensities are in parenthesis. We have collected the experimental data from these references.(Bohn, Brabson, & Andrews 1992; Dellavedova 1991; Ioannoni et al. 1989; Winnewisser & Hocking 1980)

| mode | HCOSH | *exp. gas* | *exp. argon* | *exp. xenon* | | HCSOH | | HCSSH | *exp. gas* | exp. argon |
|---|---|---|---|---|---|---|---|---|---|---|
| ***trans*** | | | | | | | | | | |
| ν1 C-H str | 2824 (45.4) | 2844 | 2867.6 | 2835.3 | ν1 O-H str | 3483 (34.4) | ν1 C-H str | 2956 (8.5) | 2964 | |
| ν2 S-H str | 2551 (2.0) | 2590 | 2597.7 | 2569.7, 2567.5 | ν2 C-H str | 3001 (11.5) | ν2 S-H str | 2537 (0.7) | | |
| ν3 C=O str | 1744 (348.5) | 1722 | 1708.3 | 1703.2 | ν3 HCOH def. + | 1417 (165.4) | ν3 CH rock | 1287 (46.0) | 1283 | 1265 |
| ν4 CH rock | 1352 (10.1) | 1340 | 1363.4, 1350.8 | 1349.7, 1348.3 | ν4 C-O str | 1229 (219.7) | ν4 C=S str | 1050 (132.6) | 1057 | 1082 |
| ν5 CH wag oop | 926 (4.5) | 932 | 930.0, 928.1 | 935.3 | ν5 HCOH def. + | 1187 (131.3) | ν5 CSH def + ( | 922 (43.8) | 936 | 926 |
| ν6 CSH def + C-S str | 918 (43.0) | 924 | 932.5 | 928.4, 925.9 | ν6 C H wag oo | 937 (40.4) | ν6 C H wag o | 834 (30.8) | 824 | 807 |
| ν7 C-S str | 647 (120.4) | 676 | 671.2, 669.3 | 684.5 | ν7 C=S str | 938 (19.0) | ν7 C-S str | 666 (43.7) | 683 | 726 |
| ν8 SCO scissors | 424 (5.2) | 431 | 429.2 | 440.6, 438.1 | ν8 COH | 656 (69.3) | ν8 CSH | 425 (5.9) | | |
| ν9 CSH | 417 (22.0) | 400 | | | ν9 SCO scissors | 459 (12.9) | ν9 SCS scisso | 320 (2.3) | | |
| 2ν1 | 5405 (1.3) | | | | | 6792 (5.0) | | 5782 (2.3) | | |
| 2ν2 | 5003 (0.3) | | | | | 5881 (1.5) | | 4975 (0.6) | | |
| 2ν3 | 3466 (7.0) | 3420 | | | | 2799 (0.2) | | 2553 (0.1) | | |
| 2ν4 | 2656 (12.1) | 2647 | | | | 2467 (0.9) | | 2092 (0.8) | | |
| 2ν5 | 1848 (2.6) | 1840 | | | | 2384 (0.4) | | 1822(0.1) | | |
| 2ν6 | 1829 (1.3) | | | | | 1868 (0.8) | | 1664 (0.9) | | |
| ν2+ν1 | 5320 (0.0) | | | | | 6482 (0.1) | | 5491 (0.0) | | |
| ν3+ν1 | 4512 (1.8) | | | | | 4886 (1.3) | | 4244 (0.8) | | |
| ν3+ν2 | 4295 (0.1) | | | | | 4413 (0.6) | | 3820 (0.1) | | |
| ν4+ν1 | 4059 (0.1) | | | | | 4710 (1.4) | | 4004 (0.2) | | |
| ν4+ν2 | 3902 (0.0) | | | | | 4248 (0.1) | | 3587 (0.0) | | |
| ν4+ν3 | 3098 (1.5) | | | | | 2638 (1.1) | | 2330 (0.7) | | |
| ν5+ν1 | 3682 (0.1) | | | | | 4671 (0.6) | | 3872 (0.1) | | |
| ν5+ν2 | 3476 (0.0) | | | | | 4187 (0.1) | | 3434 (1.0) | | |
| ν5+ν3 | 2663 (0.2) | 2620 | | | | 2587 (0.6) | | 2197 (0.0) | | |
| ν5+ν4 | 2280 (0.0) | | | | | 2424 (2.5) | | 1964 (0.1) | | |
| | | | | | | | | | | |
| ***cis*** | | | | | | | | | | |
| ν1 C-H str | 2802 (55.9) | 2842 | 2866.8 | 2833.0 | ν1 O-H str | 3565 (51.4) | ν1 C-H str | 2964 (11.9) | | |
| ν2 S-H str | 2548 (0.1) | 2590 | 2601.7 | 2578.7 | ν2 C-H str | 2951 (30.4) | ν2 S-H str | 2517 (1.8) | | |
| ν3 C=O str | 1752 (369.5) | 1722 | 1712.7 | 1705.7 | ν3 HCOH def. + | 1434 (65.0) | ν3 CH rock | 1262 (75.5) | 1256 | |
| ν4 CH rock | 1335 (17.3) | 1350 | 1342.5, 1341.0 | 1339.9, 1337.2 | ν4 HCOH def. | 1238 (6.9) | ν4 C=S str | 1074 (95.9) | 1081 | |
| ν5 CSH def + C-S str | 930 (52.4) | 949 | 946.3, 945.3 | 946.9 | ν5 C-O str | 1185 (447.8) | ν5 CSH def + ( | 909 (73.9) | | |
| ν6 CH wag oop | 916 (3.9) | 924 | 916.2 | 917.8, 915.4 | ν6 C=S str | 942 (16.0) | ν6 C H wag o | 806 (27.5) | 795 | |
| ν7 C-S str | 687 (90.9) | 718 | 715.0, 705.2 | 728.7, 719.6 | ν7 C H wag oo | 900 (10.7) | ν7 C-S str | 704 (35.4) | 710 | |
| ν8 SCO scissors | 409 (3.6) | 431 | | | ν8 COH | 499 (106.3) | ν8 CSH | 347 (12.6) | | |
| ν9 CSH | 346 (2.7) | 384 | 407.1, 404.9 | 410.3 | ν9 SCO scissors | 469 (4.3) | ν9 SCS scisso | 298 (0.1) | | |
| 2ν1 | 5463 (1.6) | | | | | 6959 (9.3) | | 5796 (2.7) | | |
| 2ν2 | 4990 (0.3) | | | | | 5711 (1.7) | | 4927 (0.8) | | |
| 2ν3 | 3481 (7.5) | 3420 | | | | 2814 (3.3) | | 2504 (0.1) | | |
| 2ν4 | 2643 (3.1) | 2675 | | | | 2469 (1.4) | | 2141 (0.7) | | |
| 2ν5 | 1851 (0.7) | 1840 | | | | 2355 (4.0) | | 1811 (0.1) | | |
| 2ν6 | 1831 (6.3) | | | | | 1880 (1.4) | | 1610 (0.9) | | |
| ν2+ν1 | 5351 (0.0) | | | | | 6487 (0.1) | | 5482 (0.0) | | |
| ν3+ν1 | 4550 (1.4) | | | | | 4980 (0.2) | | 4228 (0.6) | | |
| ν3+ν2 | 4298 (0.0) | | | | | 4330 (0.9) | | 3776 (0.1) | | |
| ν4+ν1 | 4136 (0.0) | | | | | 4795 (0.6) | | 4037 (0.1) | | |
| ν4+ν2 | 3881 (0.1) | | | | | 4151 (0.2) | | 3586 (0.2) | | |
| ν4+ν3 | 3089 (1.5) | | | | | 2640 (1.1) | | 2330 (0.7) | | |
| ν5+ν1 | 3732 (0.1) | | | | | 4751 (0.4) | | 3875 (0.0) | | |
| ν5+ν2 | 3460 (0.2) | | | | | 4111 (0.3) | | 3413 (0.2) | | |
| ν5+ν3 | 2680 (0.5) | 2647 | | | | 2597 (1.6) | | 2167 (0.3) | | |
| ν5+ν4 | 2263 (0.1) | | | | | 2410 (0.5) | | 1979 (0.3) | | |

Table 2. The effective rotational constants of three thioformic acid isomers in MHz. The computational structures were optimized with B3LYP/aug-cc-pVTZ level of theory and the experimental values are from these references: [a](Hocking & Winnewisser 1976a, 1976b) and [b](Bak, Nielsen, & Svanholt 1978; Prudenzano et al. 2018). These computational rotational constants are not accurate enough to be directly used for astronomical observations.

| **HCOSH** | | | | **HCSOH** | | **HCSSH** | | | |
|---|---|---|---|---|---|---|---|---|---|
| comp. | *exp. [a]* | comp. | *exp. [a]* | comp. | comp. | comp. | *exp. [b]* | comp. | *exp. [b]* |
| *trans* | | *cis* | | *trans* | *cis* | *trans* | | *cis* | |
| 62488.04 | *62036.09* | 63618.16 | *62927.71* | 64988.00 | 72765.87 | 50067.97 | 49206.00 | 49607.99 | 48572.40 |
| 6055.67 | *6425.31* | 6057.48 | *6134.26* | 6278.78 | 6135.84 | 3397.69 | 3447.53 | 3444.16 | 3498.75 |
| 5520.61 | *5569.64* | 5530.86 | *5584.75* | 5725.60 | 5658.67 | 3181.76 | 3219.47 | 3220.55 | 3261.42 |

Table 3. Our experimental infrared absorption frequencies of HCOSH (TFA) rotamers and their UV-photoproducts in argon and xenon matrixes.

| mode/product | argon matrix | xenon matrix |
|---|---|---|
| ***trans-TFA*** | | |
| ν1 C-H str | 2867.6 | 2835.3 |
| ν2 S-H str | 2597.7 | 2569.7, 2567.5 |
| ν3 C=O str | 1708.3 | 1703.2 |
| ν4 CH rock | 1363.4, 1350,8 | 1349.7, 1348.3 |
| ν5 CSH def + C-S str | 930.0, 928.1 | 935.3 |
| ν8 CH wag oop | 932.5 | 928.4. 925.9 |
| ν6 C-S str | 671.2. 669.3 | 684.5 |
| ν7 SCO scissors | 429.2 | 440.6, 438.1 |
| ν9 CSH | N/A | N/A |
| ***cis-TFA*** | | |
| ν1 C-H str | 2866.8 | 2833.0 |
| ν2 S-H str | 2601.7 | 2578.7 |
| ν3 C=O str | 1712.7 | 1705.7 |
| ν4 CH rock | 1342.5, 1341.0 | 1339.9, 1337.2 |
| ν5 CSH def + C-S str | 946.3, 945.3 | 946.9 |
| ν8 CH wag oop | 916.2 | 917.8, 915.4 |
| ν6 C-S str | 715.0, 705.2 | 728.7, 719.6 |
| ν7 SCO scissors | N/A | N/A |
| ν9 CSH | 407.1, 404.9 | 410.3 |
| ***Photoproducts*** | | |
| $H_2S$ | 2856.1, 2681.3 | 2673.8, 2672.1 |
| $H_2S \cdots CO$ | 2500.6, 2496.7, 2484.9, 2474.4 | N/A |
| $CO_2$ | 2345.0, 2339.1 | 2335.1, 2334.5 |
| CO | 2138.2 | 2138.0, 2137.0 |
| OCS | 2049.6 | 2046.4 |

# Appendix

Supporting information. Minimum energy and transition state structures of thioformic acid isomers in Cartesian coordinates, the units are in Ångströms (Å).

```
5
HCOSH cis
C        -0.67644        0.42727       -0.00018
H        -0.75483        1.52611        0.00005
O        -1.62154       -0.30518        0.00010
S         1.01285       -0.16875       -0.00002
H         1.58024        1.05170        0.00055
5
HCOSH trans
C         0.67989        0.43228       -0.00005
H         0.79612        1.52827       -0.00014
O         1.60169       -0.33129       -0.00033
S        -1.05305       -0.00832        0.00017
H        -0.84023       -1.33852        0.00038
5
HCOSH TS
C        -0.72542        0.42572        0.01196
H        -0.86933        1.51622        0.05522
O        -1.62350       -0.34983       -0.00513
S         1.06531       -0.06851       -0.08380
H         1.16498       -0.17573        1.25480
5
HCSOH cis
C        -0.42516        0.44724        0.00001
H        -0.66682        1.51043       -0.00009
S         1.09624       -0.09644       -0.00007
O        -1.50119       -0.35290        0.00012
H        -2.31258        0.17236        0.00022
5
HCSOH trans
C        -0.43616        0.49490        0.00011
H        -0.66119        1.55767       -0.00022
S         1.07205       -0.10671       -0.00001
O        -1.56696       -0.20835       -0.00004
H        -1.33899       -1.15287        0.00007
5
HCSOH TS
C        -0.41005        0.48067        0.01481
H        -0.60143        1.55291        0.06139
S         1.08642       -0.11734        0.00282
O        -1.54919       -0.25676       -0.11637
H        -1.92743       -0.50542        0.73561
5
HCSSH cis
C        -0.09909        0.59608        0.00006
H        -0.01104        1.67925        0.00006
S        -1.53816       -0.15218       -0.00006
S         1.43698       -0.23036        0.00003
H         2.22445        0.86486       -0.00001
```

```
5
HCSSH trans
C          0.09853         0.59833        -0.00007
H          0.03170         1.68456        -0.00022
S          1.53357        -0.16076        -0.00006
S         -1.50229        -0.08282         0.00011
H         -1.12324        -1.37718        -0.00010
5
HCSSH TS
C         -0.14752         0.61173         0.02027
H         -0.11761         1.69724         0.05221
S         -1.54397        -0.18155        -0.00231
S          1.50397        -0.13597        -0.08635
H          1.64285        -0.28726         1.24475
```